\documentstyle[psfig]{article}
\newcommand{\etal}{{\it et al.}}
\newcommand{\kms}{km~s$^{-1}$}

\newcommand{\msol}{M$_{\odot}$}

\newcommand{\pyr}{yr$^{-1}$}
\newcommand{\mum}{$\mu$m}
\newcommand{\CHp}{CH$^{+}$}
\newcommand{\philband}{$\rm A^{1}\Pi_{u}-X^{1}\Sigma^{+}_{g}$}

\newcommand{\redband}{ $\rm A^{2}\Pi    -X^{2}\Sigma^{+}$}

\begin{document}
 \title{CIRCUMSTELLAR MOLECULAR LINE ABSORPTION AND EMISSION \\
        IN THE OPTICAL SPECTRA OF POST-AGB STARS
 \thanks{Based on observations with the WHT/UES}}
 \author{Eric~J.~Bakker, Henny~J.G.L.M.~Lamers, \\
    SRON Laboratory for Space Research Netherlands, SRON-Utrecht \\
    Astronomical Institute, University of Utrecht                \\
         L.B.F.M.~Waters \\
    Astronomical Institute 'Anton Pannekoek', University of Amsterdam \\
    SRON Laboratory for Space Research Netherlands, SRON-Groningen \\
         Ton~Schoenmaker \\
    Kapteyn Sterrenwacht Roden}

 \maketitle

 \begin{abstract}
 We present a list of post-AGB stars showing
 molecular line absorption and emission in the optical spectrum.
 Two objects show \CHp, one in emission and one
 in absorption, and ten stars show C$_{2}$ and CN in absorption.
 The Doppler velocities of the C$_{2}$ lines and the rotational
 temperatures indicate that the line forming region is the AGB remnant.
 An analysis of the post-AGB  stars of which CO millimeter data is
 available suggests that the C$_{2}$ expansion velocity is of the same order
 as the CO expansion velocity. HD~56126 has been studied in detail and
 we find a mass-loss rate of $\dot M=2.8\times10^{-4}$ \msol \pyr,
 $f_{\rm C_{2}}=2.4 \times10^{-8}$ and
 $f_{\rm CN}=1.3\times10^{-8}$.  The mass loss derived from C$_{2}$ is
 significantly larger than $\dot M=1.2\times10^{-5}$ \msol \pyr~ derived from
  CO.
 We find that all objects  with the 21\mum~ feature in emission
 show C$_{2}$ and CN absorption, but not all objects with C$_{2}$
 and CN detections show
 a 21\mum~ feature.

 \noindent
 {\bf keywords: } molecules --  physical conditions in AGB remnant  --
 mass-loss history on AGB

 \end{abstract}

 \section{Introduction}

\begin{figure*}
 \centerline{\hbox{\psfig{figure=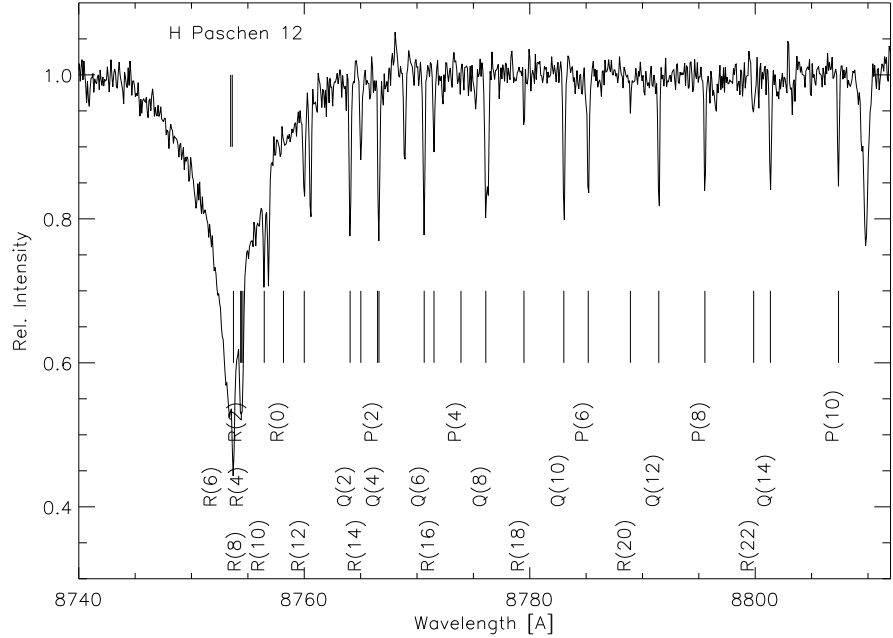,width=\columnwidth}}}
 \caption{The Phillips (2,0) band in the spectrum of HD~56126. The
            C$_{2}$ absorption lines are not resolved.}
 \label{edinfig-phillips20}
 \end{figure*}

\begin{figure*}
 \centerline{\hbox{\psfig{figure=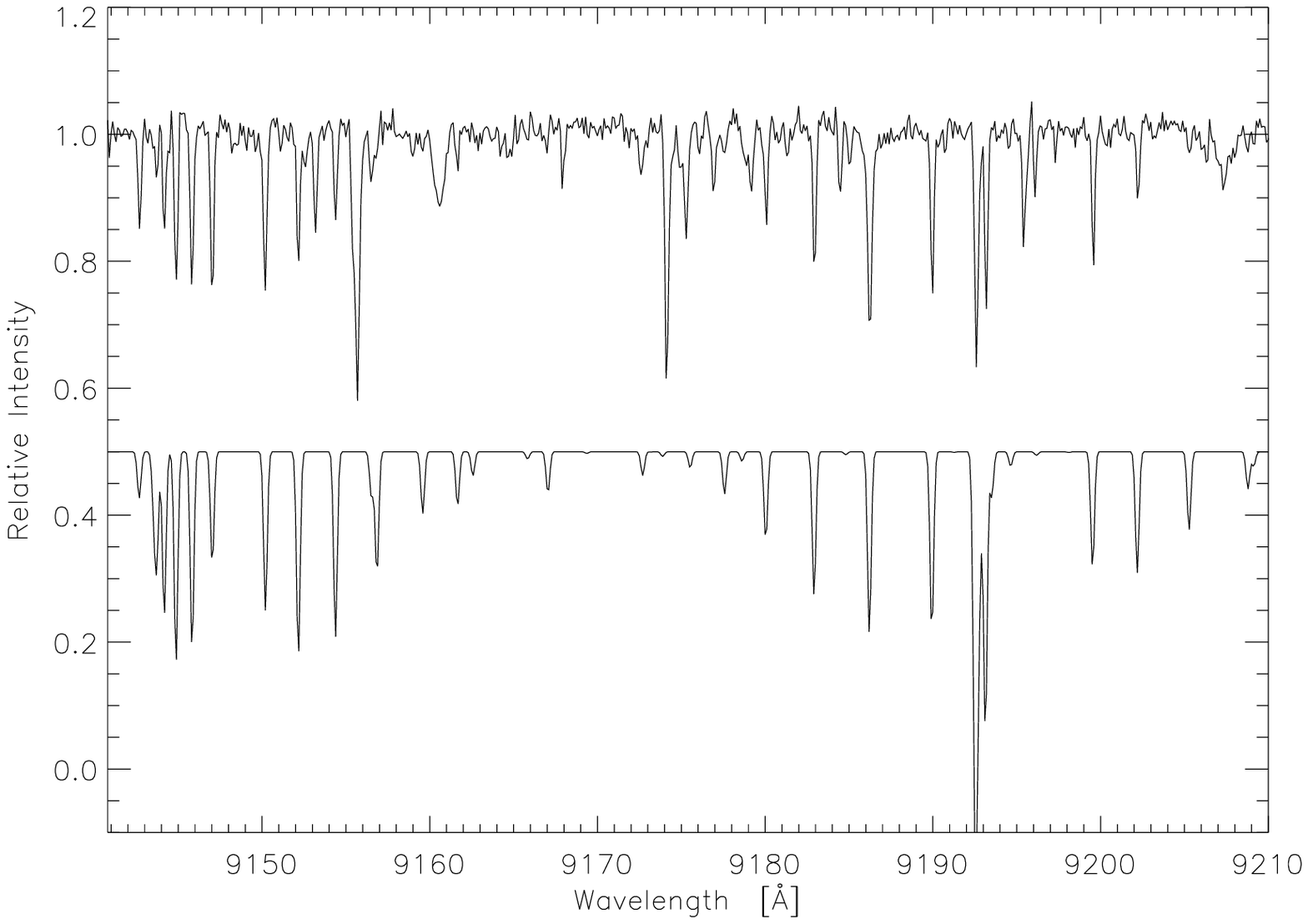,width=\columnwidth}}}
 \caption{The CN \redband~ (1,0) band in the spectrum of HD~56126.
 The computed synthetic spectrum is shifted down by 0.5.}
 \label{edinfig-cn10}
 \end{figure*}

 The study of molecular lines in the optical spectra of post-AGB stars
 started with the paper by Waelkens \etal~ (1992) in
 which they presented a band of narrow emission lines in
 the spectrum of the Red Rectangle (HD~44179). These emission
 lines were identified as the (0,0) C$_{3}$
 band of \CHp~ by Balm and Jura (1992).
 The presence
 of the Swan and Phillips bands  of C$_{2}$ (Fig.~\ref{edinfig-phillips20})
 and red system bands
 of CN (Fig.~\ref{edinfig-cn10})
 in the spectrum of HD~56126   studied by Bakker
 (1994), showed that the outflow velocity and the
 excitation conditions of C$_{2}$ suggest that these lines are formed in the
 AGB remnant. This opens new possibilities
 to study the physical conditions in AGB remnants. Hrivnak
 (1995) presents a list of post-AGB stars showing C$_{2}$ and C$_{3}$
 in absorption.
 Here we present  the observations of post-AGB stars
 that show molecular lines in the optical spectrum.
 For several stars the excitation temperature and  outflow velocity
 have been determined. The C$_{2}$ absorption in HD~56126
 is studied in detail by modeling the population density
 over the rotational quantum number (Section~\ref{edinsec-hd56126}).

 \section{Detection of Molecules}

In Table \ref{edintab-vel} we list the first results of our work.
For each star the 3th and 4th column give the expansion velocity
derived from
the optical (C$_{2}$,CN,\CHp) and the CO millimeter or OH maser line
respectively.
The 5th and 6th column give the rotational temperature and column density
derived from C$_{2}$ (3,0) or \CHp (0,0) band.
Negative velocities are due to outflow of material.
The C$_{2}$ line absorption is formed closer to the star than the CO line
emission, the two different molecules trace different material and hence
different stages of the AGB evolution. Our data may indicate that the
expansion velocity decreases as the star evolves along the AGB
(Table~\ref{edintab-vel}).  However a larger sample is needed to quantify this
 result. Stars showing the
 unidentified 21\mum~ emission feature all exhibit C$_{2}$ and CN  absorption,
 strongly suggesting that the 21\mum~ feature is from carbon-rich material,
 but not all stars with C$_{2}$ absorption show the 21\mum~ feature.

\begin{table}
\caption{Expansion velocities of the AGB remnant}
\label{edintab-vel}
\begin{tabular}{llrrrrl}
 Object         &Id$^{*}$&$\delta v_{\rm optical}$
                                    &$\delta v^{+}$
                                           &$T_{rm rot}$&$N_{\rm
Boltz}\times10^{14}$&Remark            \\
                &        &\multicolumn{2}{c}{[\kms]} &[K]&[$\rm cm^{-2}$] &\\
IRAS~04296+3429 & C$_{2}$  &$-7.5$     &$-12  $ &  138    &14&
\\
IRAS~05113+1347 & C$_{2}$  &$-5.2$     & no     &  198    &12&not observed in
CO\\
HD~56126        & C$_{2}$  &$-8.8$     &$-10.0$ &  240    &20&
\\
IRAS~08005-2356 & C$_{2}$  &$38.1$     &$-50.0$ &  150    &31&OH maser
\\
IRC~+10216      & C$_{2}$  &$17.9$     &$-15.7$ &         &  &AGB star
\\
AFGL~2688       & C$_{2}$  &$15.0$     &$-22.8$ &   56    &26&
\\
HD~44179        & \CHp   &$-4.5$ em. &$-3   $ &  267    & 2&
\\
HD~213985       & \CHp   &$-5.4$ abs.& nd   &           &  &not detected in
CO\\
\end{tabular}
\newline
$^{*}$ all stars with C$_{2}$ do also show CN absorption \newline
$^{+}$ from CO millimeter emission. For IRAS~08005-2356 there is no CO data
available
and we used the velocities from the OH maser line
\end{table}

 \section{Modeling the C$_{2}$ absorption Bands of HD~56126}
 \label{edinsec-hd56126}

\begin{figure*}
 \centerline{\hbox{\psfig{figure=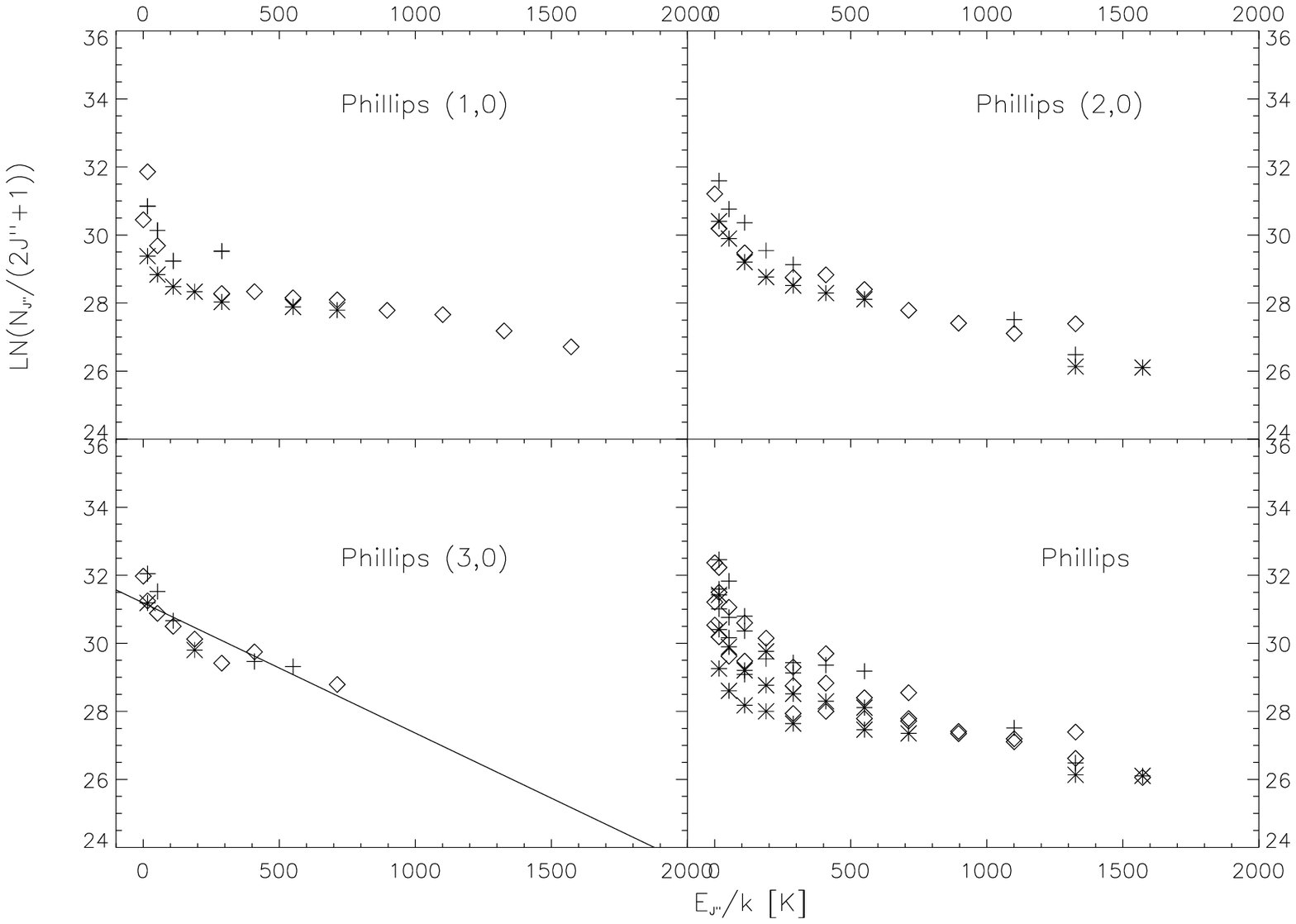,width=\columnwidth}}}
 Rotational Diagram of the C$_{2}$ \philband~
 \caption{ (1,0), (2,0),  and (3,0) absorption  band.
            The P, Q, and R branches are denoted by a
            plus, asterisk, and a square respectively. Only the weakest
            rotational band (3,0) is optically thin and
            gives an almost linear relation in the rotational diagram
            with $T_{\rm rot}=240$~K and
            $N_{\rm Boltz}= 20\times10^{14}$~cm$^{-2}$.}
 \label{edinfig-rotc2}
 \end{figure*}

 \begin{figure*}
 \centerline{\hbox{\psfig{figure=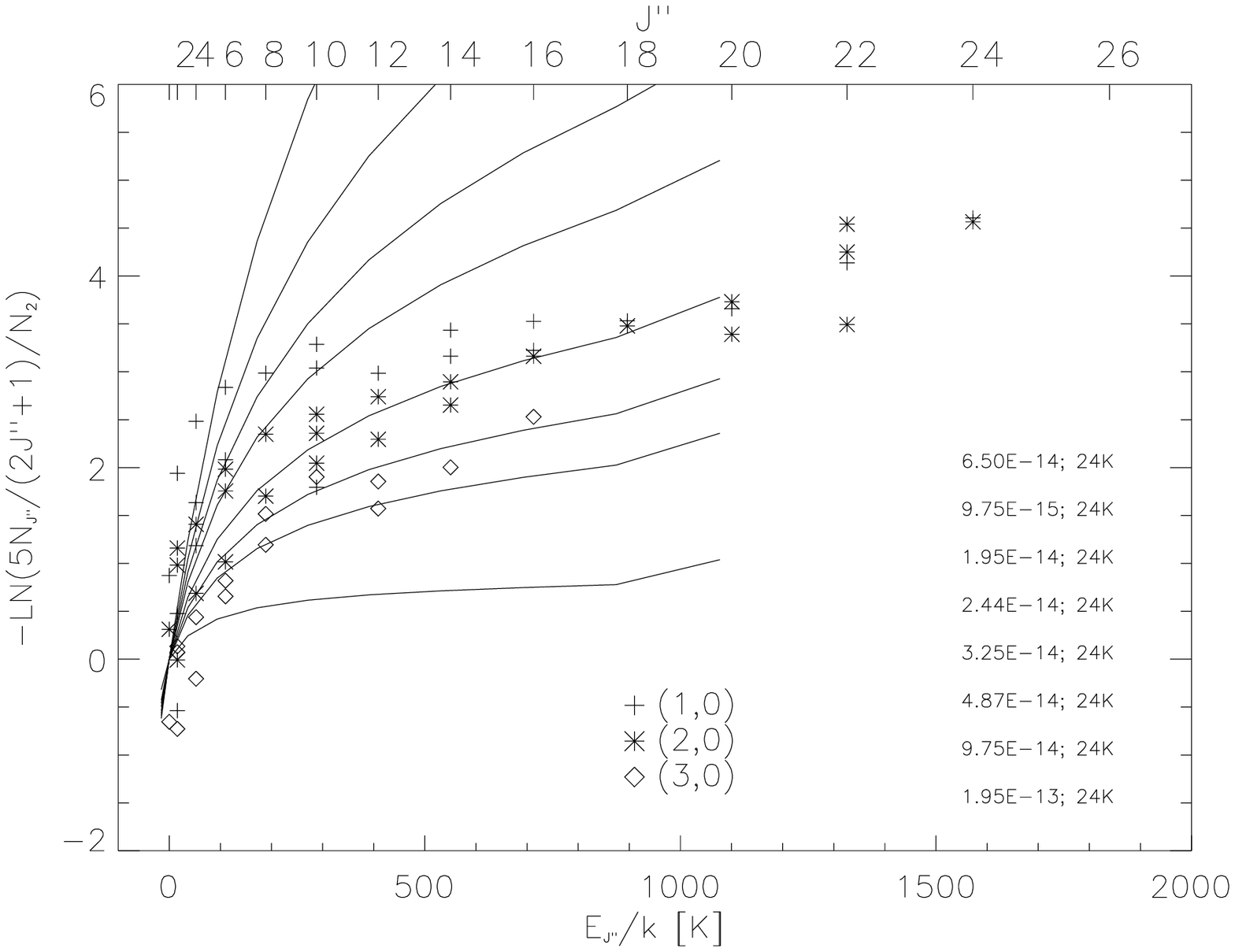,width=\columnwidth}}}
 \caption{Relative rotational diagram of the C$_{2}$ \philband~
 (1,0), (2,0), and (3,0) absorption band of HD~56126
 with the model fit for $T_{\rm kin}= 24$~K superimposed.}
 \label{edinfig-relrotc2}
 \end{figure*}

 From the rotational diagram of the C$_{2}$ bands
 (Fig.~\ref{edinfig-rotc2}) we find $T_{\rm rot}=240$ K
and for the CN (1,0) band
 we find $T_{\rm rot}=24$ K with a column density of
 $N_{\rm C_{2}} = 20\times10^{14}$ cm$^{-2}$ and
 $N_{\rm CN} = 11\times10^{14}$ cm$^{-2}$.
 Unlike CN, C$_{2}$ is a homo-nuclear molecule having no permanent
  dipole moment and the
 rotation temperature is not a good indicator of the kinetic temperature
 because of pumping by the stellar radiation field.
 We use the CN rotational temperature
 as a measure of the kinetic temperature of the gas.

 We have tried to fit the relative rotational diagram
 (Fig.~\ref{edinfig-relrotc2})
 with the population
 density distribution derived from modeling the C$_{2}$ excitation by
 taking into account radiative pumping and collisional de-excitation
 (Van Dishoeck and Black 1982).
 The best fit is reached for
 $ n_{c} \sigma /I = 3.25\times10^{-14}$. Taking the  latest H$_{2}$-$C_{2}$
cross section
 of $\sigma = 7.8\times10^{-16}$ cm$^{-2}$ (Phillips 1994)
 the mass-loss rate can be calculated using Eq.~\ref{edineq-mdotmodel}.
 This gives a mass-loss rate of $\dot M_{\rm C_{2}}=2.8\times10^{-4}$ \msol
\pyr.
 \begin{equation}
 \label{edineq-mdotmodel}
 \dot M = 2.8\times 10^{-4}  .
                   \left(
                   \frac{v_{exp}}{8.8} .
                   \frac{n_{c}}{1.7\times 10^{7}} .
                   \frac{\sigma}{7.8 \times10^{-16}} .
                   \frac{4.1\times 10^{5}}{I}
                   \right)
                   ~M_{\odot}~ {\rm yr^{-1}}
 \end{equation}
 Fitting an optically thin dust model (Waters \etal~ 1988) to the
 the spectral energy distribution  of HD~56126
  yields a dust inner radius of $2\times10^{3} R_{*}$
 and $\dot M =1.9\times10^{-4}$
 \msol \pyr. Assuming that the molecular line
 absorption originates at this dust inner radius we find
 a particle abundance relative to H$_{2}$ of $f_{\rm C_{2}}=2.4\times10^{-8}$
and
 $f_{\rm CN}=1.3\times10^{-8}$.
 Table~\ref{edintab-mdot} summarizes the mass-loss rates derived for HD~56126,
 where we have scaled the different mass-loss rates to
 $R_{*}=50R_{\odot}$, $T_{\rm eff}=6500$ K ($\log L=3.6$) and $D=2.7$ kpc.
The mass-loss rate derived from C$_{2}$ and from the IR excess are
significantly higher than that derived from the CO
millimeter emission, which might indicate that the mass-loss rate increased
dramatically towards the end of the AGB.

\begin{table}
\caption{Overview of derived mass-loss rates for HD~56126}
\label{edintab-mdot}
\begin{tabular}{lll}
Modeling C$_{2}$ excitation  &$2.8\times10^{-4}$ \msol \pyr & this study
                 \\
Infrared excess from dust    &$1.9\times10^{-4}$ \msol \pyr & this study;
gas/dust=100           \\
CO millimeter emission       &$1.2\times10^{-5}$ \msol \pyr & $^{12}$CO(2-1)
Omont \etal~ 1993   \\
\end{tabular}
\end{table}

\section*{Acknowledgements}

The authors want to thank Ewine van Dishoeck and Christoffel Waelkens
for the stimulating and constructive discussions on this work.
EJB was supported by grant no. 782-371-040 by ASTRON,
which receives funds from NWO. LBFMW acknowledges financial
support from the Royal Dutch Academy of Arts and Sciences.
This research has made use of the Simbad database, operated at
CDS, Strasbourg, France.

\end{document}